\documentstyle[twocolumn,prl,aps,epsf]{revtex}
\newcommand{\epsplace}[1]{\epsfxsize=3.3in \centerline{\epsfbox{#1}}}
\begin{document}

\title{Unexpected scenario of glass transition in polymer globules: an exactly enumerable model}

\author{Rose Du,${}^{1}$ Alexander Yu. Grosberg,${}^{1,2}$
Toyoichi Tanaka,${}^{1}$ and Michael Rubinstein${}^{3}$}

\address{
${}^1$Department of Physics and Center for Materials Science and
Engineering, \\ Massachusetts Institute of Technology, Cambridge,
Massachusetts 02139,  USA \\
${}^2${\em On leave from:\/} Institute of Chemical Physics,
Russian Academy of Sciences, Moscow 117977, Russia\\
${}^{3}$Department of Chemistry, University of North Carolina,
Chapel Hill, North Carolina 27599-3290, USA
}
\address{ {\em \bigskip \begin{quote}
We introduce a lattice model of glass transition
in polymer globules.  This model exhibits a 
novel scenario of ergodicity breaking in which the 
disjoint regions of phase space do not arise uniformly,
but as small chambers whose number increases exponentially 
with polymer density.  Chamber sizes obey power law distribution, 
making phase space similar to a fractal foam.  This clearly
demonstrates the importance of the phase space geometry and topology 
in describing any glass-forming system, such as semicompact polymers 
during protein folding.
\end{quote} }}

\maketitle

Although it is generally agreed that the most fundamental  
feature common to a large variety of different glasses is some form of ergodicity 
breaking, this phenomenon is still poorly understood \cite{Angell95}.   
The common scenario implies that upon lowering the temperature, increasing the pressure, or 
otherwise suppressing the thermal agitation, phase space is effectively divided  
into domains, then each of the domains is subdivided into smaller domains, and 
the process continues in an ultrametric (tree-like) fashion.  
 
One problem with the theoretical understanding of glasses stems from the lack 
of solvable models in the field.  In the 
present paper, we suggest and explore one such model in which ergodicity is 
explicitly broken.  To our surprise, this 
model exhibits a scenario of glass transition which is dramatically 
different from the above mentioned classical picture.  Specifically, 
phase space of our model appears to consist of one huge valley and a 
large number of very small ``chambers.'' 
When the system is further restricted, the 
number of small chambers grows at the expense of the single big valley, 
but they remain small in the sense that the distribution of domain sizes 
remains bimodal, with enormous gap between the 
big valley and small chambers.  However the small
chambers are themselves of different sizes, and their 
distribution appears to follow a power law, thus indicating fractal 
nature of the disjoint phase space.     

The motivation for our model comes from the works on the theory of protein 
folding, where a toy protein is usually presented 
as a cubic lattice polymer with a quenched sequence of monomer species.  
There is a sophisticated theory 
\cite{IIM,pande98b} which predicts the equilibrium 
freezing transition for maximally compact heteropolymers at some temperature $T_f$.   
Below $T_f$, only very few of the compact conformations contribute to the 
partition function, although exponentially many of them contribute above the 
transition.   It is natural to assume that $(T=T_f , \phi = \phi_{\max})$ is the 
end point of some freezing phase transition line on the temperature $T$ vs. density $\phi$ 
phase diagram (see also \cite{plotkin96,pande97b}).  To gain more insight into this phase diagram one 
may want to consider the  opposite extreme of very high temperature, 
and varying density, $0 < \phi \leq \phi_{\max}$.  At high temperature 
all interactions are irrelevant except for the excluded volume one, and we have 
a homopolymer.  
Thus, we may expect some kind 
of homopolymer glass transition to occur along the $1/T = 0$ line, at a
certain density $\phi_g$ (see also \cite{Onuchic}).   
Although the relation 
of this transition to the freezing one at 
$\phi = \phi_{\max}$ is not clear,
and the question of a glass transition in the protein folding context 
has recently been the subject of a debate \cite{wolynesglass,shaknoglass},  
we think that the glass transition in a homopolymer globule deserves 
attention and our goal in the present paper is to  study a model of this phenomenon. 

We shall operate with the $3 \times 3 \times 3$ segment of the cubic lattice, 
which is a standard tool in protein folding studies.   In order to address 
the gradual change of density, we shall 
change (by steps of one) the length $N$ of polymer chains confined 
within the 27-cube.  When ergodicity is in question, 
the rules of dynamics are to be considered as an integral part of the model.  
We use the conventional set of elementary moves 
employed in Monte Carlo simulations of polymers, including 
end flips, corner flips, and crankshaft moves. 
As soon as the model is specified, including both 
the set of conformations and the set of elementary moves, 
the conformational space should be viewed as a graph, in which nodes 
represent conformations, and edges represent possible moves 
transforming one conformation into another.  The question of ergodicity 
is now reduced to the one of graph connectivity.  If the graph of 
conformations contains only one connected component, the system is 
ergodic: every conformation can be transformed into any 
other conformation by a sequence of allowed elementary moves.   
On the other hand, if the graph of conformations 
consists of two or more disconnected components, ergodicity is 
obviously broken.  

We begin with the exhaustive enumeration of  
all self-avoinding $N$-mers, $1 < N \leq 27$,
that are confined within the 27-cube.  
As indicated in Table 1, the number 
of conformations, $\Omega_N$, is peaked at 
$N=22$:  at smaller $N$, each extra monomer adds to the degrees of freedom, 
at larger $N$ the restrictions due to the confinement become increasingly 
severe.  Before proceeding to examine the phase space
in detail, one should note that the move set has a property
that if the lattice sites were colored black and
white in an alternating manner, the monomers could only move between
sites of the same color.  This immediately results in the
splitting of the phase space into two regions,  but   
this is unrelated to the ergodicity breaking. 

\noindent {\bf Table 1:} {\sf Enumeration of polymer chains of length $N$ within 
a $3\times3\times3$ cube.  
``Small chambers'' refer to all domains but 
the largest valleys.  
Data for $N$ from 5 through 13 (for which the numbers of conformations, $\Omega_N$, 
are equal, respectively, 17, 58, 193, 625, 1884, 5445, 14332, 36208, 8182) are not shown,  
because there are no small chambers.}
\begin{tabular}{llrrr}\label{tab:enumbox}\\
\tableline \tableline
chain & volume& \# of & \# of small   & \# of confs in \\ 
length    & fraction & confs & chambers & small chambers \\ 
$N$ & $\phi=N/27$ & $\Omega_N$ & $\sum n_m$ & $\sum m \cdot n_m$ \\  \tableline
14&0.52 & 177018 & 0 & 0\\
15&0.56 &  337118& 4 & 4\\
16&0.59 & 632078 & 18 & 28\\
17&0.63 & 1018452 & 54 & 112\\
18&0.67 & 1633622 & 142 & 320\\
19&0.70 & 2199836 & 742 & 1324\\
20&0.74 & 2964500 & 1742 & 3656\\
21&0.78 & 3226280 & 6385 & 14037\\
22&0.81& 3505858 & 12722 & 31948\\
23&0.85& 2865534 & 35507 & 108951\\
24&0.89& 2303244 & 58728 & 199158\\
25&0.93& 1199908 & 128271 & 673292\\
26&0.96 & 564368 & 147036 & 564368\\
27&1.00 & 103346 & 103346 & 103346\\
\tableline \tableline
\end{tabular}

To understand the breaking of ergodicity,
we enumerate all conformations which can be reached
from each initial conformation given the move set described above.
Note that while these moves are most frequently used in Monte Carlo simulations, 
our procedure is
an exact enumeration, not a Monte Carlo simulation.  The enumeration
is carried out by labeling every conformation so that we not only know 
how large the phase space of a given initial conformation is, but
also which conformations belong to the same domain.  To allow
for rapid searches (while checking if a conformation has already
been accounted for in a domain),  we store the labeled conformations
in a binary tree \cite{Lerman93}.  
To construct the binary tree,
note that conformations can be denoted as a sequence
of directions.  Thus we can denote an $N$-mer by an ($N$-1)-digit number.
A conformation whose ($N$-1)-digit representation is greater (less) 
than another is considered to be ``greater''(``less'') 
for the purposes of this tree.  The
left and right descendants of a node of the tree are ``less'' than and
``greater'' than the node, respectively.   To further reduce computational
time associated with searching the binary tree, we first perform
this procedure on a conformation that resides in a big chamber and
eliminate all of those conformations from the list of completely enumerated 
conformations and examine only the
remaining conformations.  

\begin{figure}
\epsplace{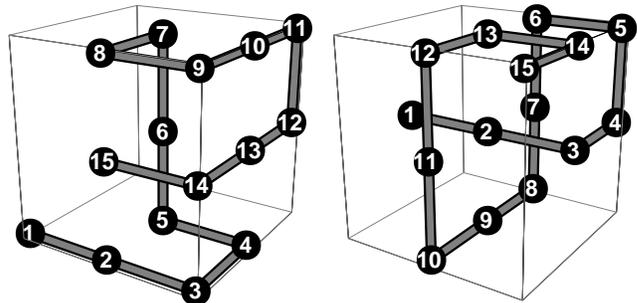}
\caption{Two 15-mer conformations which cannot
transform into any other conformation within a
$3\times3\times3$ box.
The other two such conformations are obtained by relabeling the 
monomers in the opposite order.}\label{fig:15mers}
\end{figure} 

\begin{figure}
\epsfxsize=3.0 in 
\centerline{\epsfbox{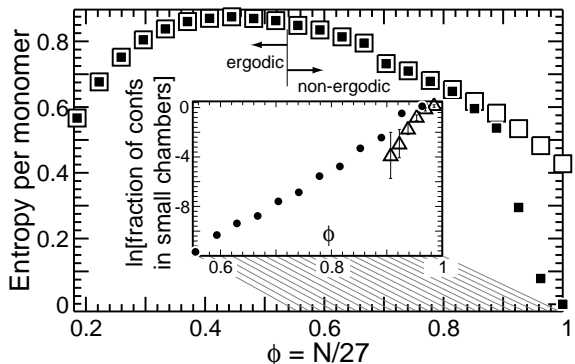}}
\caption{
The average entropy per monomer in the presence of 
barriers between chambers (black squares) vs. what
the entropy would be in the absence of such barriers
(white squares). {\bf Inset:} 
The exact data on fraction of conformations that belong to small chambers (that is, 
do not belong to the largest valley) 
for conformations in a $3\times3\times3$ cube ($\bullet$).  This fraction increases 
exponentially with polymer density.   Similar data for a 
$4\times4\times4$ cube ($\triangle$) have a $10\%$ error.  
For this figure, chambers in the 64-cube were considered
small if they have fewer than 50000 states.
}
\label{fig:spermon_fracglass}
\end{figure}

At $N=14$, the system is completely ergodic, that is, all
conformations are connected and form one big valley
(allowing for translational symmetry).  
At $N=15$, all conformations belong to the same valley 
except for four symmetrically unrelated conformations
(see Figure \ref{fig:15mers}).  It can be checked directly 
that these four conformations are unable to convert into any other conformation.
Thus ergodicity is explicitly broken, albeit by only 4 
conformations.  Similarly, at $N=16$, all conformations form one huge 
connected valley in phase space except for 28 
conformations residing in 18 small separated chambers.
This set of ``glassy'' conformations are either obtained from the
ones at the 15-mers level by appending a monomer to the head
or tail, or are very similar to them.  
This pattern continues as we 
go to larger $N$ \cite{Rose_thesis}.  Specifically, there are two classes of domains in the phase space, 
one large valley (splitted in two because of the odd-even effect) and small chambers.  The size 
separation between the two classes is enormous, making the distinction between valleys and chambers 
completely unambiguous.  Enumeration data for small chambers are presented in Table 1, where 
$m$ is the number of conformations in one chamber, and $n_m$ is 
the number of chambers of the size $m$; accordingly, $mn_m$ is the 
number of conformations in all chambers of the size $m$. 
The fraction of conformations residing in small chambers grow
exponentially with $N$ (see inset in the Figure \ref{fig:spermon_fracglass}) 
until at $N=26$, the distribution of 
domain sizes is no longer bimodal, and there is no longer
a big valley.  
Note that this picture of glass transition is qualitatively different 
from the commonly accepted one.  Ergodicity is not broken
uniformly.  Instead, the phase space consists of one
large valley containing a majority of the conformations
while tiny portions of the phase space
are pinched off at an exponential rate until the entire
phase space consists of small chambers.

Another way to look at our findings is to consider 
polymer entropy.  It can be defined in two ways.  If a chain is  
locked within one chamber, its entropy is given by $\ln m$.  Upon 
averaging over conformations, this yields    
$\left. \left[ \sum_{conf} \ln m \right] \right/ \Omega_N$, where 
$m$ for every 
conformation is the size of the chamber to which this conformation belongs.  
Alternatively, 
if there were no 
barriers between chambers, the entropy would be $\ln \Omega_N$.  The data for both 
entropies (taken per one monomer) are shown in Figure~\ref{fig:spermon_fracglass}.   
Note that while the true breaking
of ergodicity occurs at $N=15$ ($\phi \approx 0.55$), the difference between the two
entropies does not become evident until $N=23$ ($\phi \approx 0.85$).  This could explain
the difficulties encountered in determining the glass transition; while 
there exists a true transition, the
changes are too small to be measured experimentally.

\begin{figure}
\epsplace{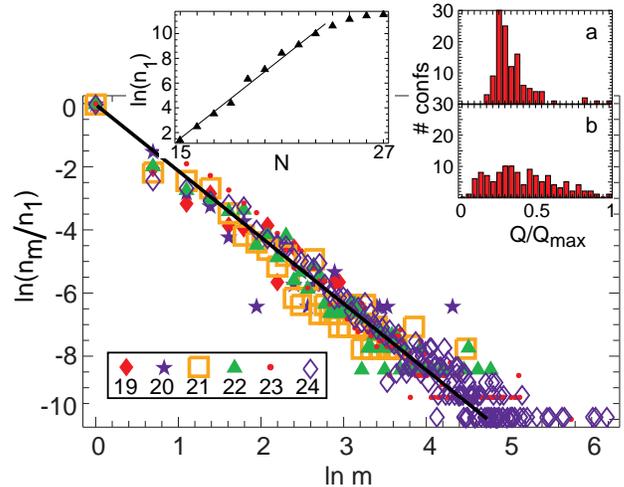}
\caption{
The normalized number of chambers, $n_m/n_1$, decays as a power 
of chamber size, $m$.  The slope of the line is $\tau = -2.15$.  
The different symbols correspond to $N$ from 19 to 24.  
{\bf Left inset:}  The number of clusters containing 
only one conformation, $n_1$,  increases exponentially with $N$ (or density 
$\phi$).  The slope of the line is $1.07$.        
{\bf Right inset:} 
The overlap distributions for two given conformations ($N=25$, $Q_{max}=24$) in a typical chamber
with  all other conformations within the same chamber.  Shown here
are the narrowest (a) and broadest (b) distributions from a chamber of size 119.} 
\label{fig:fractal}
\end{figure}

The phase space in a glassy (non-ergodic) state appears to have a
self-similar structure.  As demonstrated by figure \ref{fig:fractal}, 
the number of chambers, $n_m$, has approximately a power
law dependence on the number of conformations, $m$, in each of them, 
$n_m/n_1 \simeq m^{-\tau}$, where power $\tau$ remains between 2.1 and 2.2 
for all $N$ from 19 to 24 (at smaller $N$, there are too few points 
to claim a power law).  We found also that
the number of the smallest chambers (with $m=1$) grows
exponentially with $N$ as soon as $N$ exceeds the threshold value of 15: 
$n_1 \approx e^{1.07N-14.53}$, as shown in the left inset 
in figure \ref{fig:fractal}.   

In order to understand the structure of each chamber, we can examine
the distributions of overlaps between a given conformation in the
chamber and all other conformations within the same chamber, where
the overlap of two conformations, $Q$,  is the number
of monomer-monomer bonds these two conformations have in common. 
We find that the distribution of $Q$ is highly dependent upon the
conformation with which the overlap
is taken (see the right inset in the Figure~\ref{fig:fractal}). 
We can infer a consistent
picture of the phase space chamber as having  
small fingers extending from the main ball-like region.
Conformations which reside in these fingers have small
overlaps with most other conformations which reside in the
main ball.  As the chamber size is increased, these fingers
lengthen, so that the peak overlap decreases. 

As a preliminary test of a larger system, we examined random samples 
of 100 conformations of $N$-mers for every $N$ from 58 to 63 within 
a $4 \times 4\times4$ box and found a behavior similar 
to that of the $3 \times 3 \times 3$ system (see inset in the Figure 
\ref{fig:spermon_fracglass}). 

Switching from solid results to speculations, it is tempting  
to view the first breaking of ergodicity at $N=15$ as merely 
a signal of approaching to a real critical point at around 
$N=25$.  We can conjecture that for a very large system, in 
thermodynamic limit, there is a critical density at which 
the infinite phase space disintegrates into infinite set of 
finite disjoint domains.  We can further interpret this 
criticality in terms of some kind of percolation in the phase 
space, in this case small chambers and big valley being, 
respectively, finite and infinite clusters.  Our finding of 
self-similar fractal distribution of clusters, or 
small chambers, supports this idea qualitatively.  However, on 
the quantitative level, the interpretation of the exponent
$\tau \approx 2.1 \ {\rm to} \ 2.2$ requires better understanding of the
underlying conformational space.

Another way to look at the conjectured phase transition would be 
to examine the densities higher than critical.  We can consider 
a $(k^3-1)$-mer confined in a 
$k \times k \times k$ box, $k \gg 1$, with only one empty lattice 
node - a vacancy.  Each allowed elementary chain move can now be 
interpreted as the vacancy move in real space.  Clearly, the 
vacancy can move if and only if the conformation has the properly 
positioned corner next to it.  For a random conformation, this 
happens with some probaility less than unity.  Therefore, the 
probability that a vacancy can make $r$ consequtive steps 
decays exponentially with $r$.  Thus,  on average, 
vacancy can travel only some finite distance in real space 
\cite{distance}.   Of course, there are some zigzag conformations allowing 
vacancy to move arbitrarily far, but they are exponentially rare.      
When translated into the language of the phase space, this 
immediately explains that phase space chambers are small, in the 
sense that their sizes remain finite and do not depend on the 
overall volume of the system, even in the thermodynamic limit 
($k \to \infty$).  
The situation becomes more complicated when several vacancies 
are present, since they interact.  It would be interesting to look at this 
situation both analytically and numericaly.  
Unfortunately, 27-cube is too small for that, while computational analysis of 
clusters for 64-mer is very demanding.  We leave 
it as an unresolved challenge to understand the nature of this 
transition.  

To conclude, we speculate on the implications of our model 
for the protein folding studies.  
Numerous works \cite{Thir97} 
suggest that folding begins with a fast
collapse to a semi-compact and non-specific
globule state. In terms of our model it is natural to assume that 
such a burst collapse will bring the chain into 
one of the disjoint domains of its conformation space 
(most likely, into 
the big valley, if there remains one at a given semi-compact density).  
Speed and reliability of the further folding depends then on 
whether native state belongs to the same phase space domain or not. 
Furthermore, the observable appearance 
of the freezing transition  may be spectacularly modified by the 
foam-like structure of the phase space.  For instance, if 
denatured molten globule is restricted in one of the small 
chambers, and if the native conformation corresponds to a broad 
monomodal distribution 
of overlaps (see inset (b) in the Figure \ref{fig:fractal}), then the 
freezing behavior will be 
different from the conventional Random Energy Model behavior \cite{Eugene2D}.   

Looking from a slightly different perspective, we can view our model as
stressing the role of the underlying geometry for compact biopolymers 
and for other glass-forming systems.  In this sense, frozen conformations 
in our model 
are somewhat similar in spirit to the so-called arching 
configurations in systems of hard particles such as sand.   
Understanding this geometry presents a
significant challenge, as is convincingly proven by the example of Kepler's
conjecture on the most compact packing of hard spheres: it
took over 300 years to prove \cite{kepler}.   
However complicated the phase space geometry may be,
it is of primary importance.  If and when phase space consists of disjoint
regions, or  if regions are joined via narrow passes, 
this must be taken into account in any approach describing system 
in terms of a folding funnel \cite{BOSW95} or a reaction
coordinate \cite{Rxncoord}, or considering glass as
a separate thermodynamic phase \cite{parisiglass}.

To summarize, we have demonstrated a novel scenario for glass 
transitions using a lattice model for a homopolymer.  Unlike 
the typical picture in which
the phase space is more or less evenly divided,
the glass transition  in our model occurs via the exponential
increase of the number of small disjoint regions of phase space.
As these regions comprise an extremely small fraction of the entire
phase space, their initial appearance is hardly noticeable experimentally.  
This may contribute to the difficulty often seen in determining 
glass transition.  While these results are
highly dependent upon the restricted model which we employed
and it is unclear how common this scenario is among other
systems, it presents {\em a possiblity}.

We thank M.Kardar for the usefull discussion.  The work was supported by NSF grant DMR-9616791.


\begin{thebibliography}{99}
\bibitem{distance} Considering all corners around the vacancy independent, we found 
that vacancy is movable with probability 
$p \approx 1- \left(1-2 \frac{1}{6}\frac{1}{5} \right)^{12} \approx 0.56$, 
which corresponds to the average distance of 
$p/(1-p)^2 \approx 2.95$
\end{thebibliography}
\end{document}